\documentclass[superscriptaddress,amsmath,amssymb,aps,twocolumn,showpacs,pra]{revtex4-2}
\usepackage{graphicx,dcolumn,bm,mathrsfs,hyperref,times}

\begin{document}
\title{Synchronization and phase shaping of single photons with high-efficiency quantum memory}
\thanks{This work was supported by the Key-Area Research and Development Program of GuangDong Province (Grant No. 2019B030330001), the Key Project of Science and Technology of Guangzhou (No. 2019050001), the National Key Research and Development Program of China (Grants No. 2016YFA0302800, No. 2016YFA0301800 and 2020YFA0309500), and the National Natural Science Foundation of China (Grants No. 11822403, No. 62005082, No. 12004120, No. U20A2074, and No. U1801661), the Natural Science Foundation of Guangdong Province (Grant No. 2018A0303130066), the China Postdoctoral Science Foundation (Grant No. 2020M672681), and the GuangDong Basic and Applied Basic Research Foundation (Grant No. 2020A1515110848). }

\author{Keyu Su}
\thanks{These authors contributed equally}

\author{Yunfei Wang}\email{yunfeiwang2014@126.com}
\thanks{These authors contributed equally}
\affiliation{Guangdong Provincial Key Laboratory of Quantum Engineering and Quantum Materials, School of Physics and Telecommunication Engineering, South China Normal University, Guangzhou 510006, China}

\author{Shanchao Zhang}
\thanks{These authors contributed equally}
\affiliation{Guangdong Provincial Key Laboratory of Quantum Engineering and Quantum Materials, School of Physics and Telecommunication Engineering, South China Normal University, Guangzhou 510006, China}
\affiliation{Guangdong-Hong Kong Joint Laboratory of Quantum Matter, Frontier Research Institute for Physics, South China Normal University, Guangzhou 510006, China}

\author{Zhuoping Kong}
\author{Yi Zhong}
\author{Jianfeng Li}
\affiliation{Guangdong Provincial Key Laboratory of Quantum Engineering and Quantum Materials, School of Physics and Telecommunication Engineering, South China Normal University, Guangzhou 510006, China}

\author{Hui Yan}\email{yanhui@scnu.edu.cn}
\affiliation{Guangdong Provincial Key Laboratory of Quantum Engineering and Quantum Materials, School of Physics and Telecommunication Engineering, South China Normal University, Guangzhou 510006, China}
\affiliation{Guangdong-Hong Kong Joint Laboratory of Quantum Matter, Frontier Research Institute for Physics, South China Normal University, Guangzhou 510006, China}
\affiliation{Guangdong Provincial Engineering Technology Research Center for Quantum Precision Measurement, South China Normal University, Guangzhou 510006, China}

\author{Shi-Liang Zhu}\email{slzhu@scnu.edu.cn}
\affiliation{Guangdong Provincial Key Laboratory of Quantum Engineering and Quantum Materials, School of Physics and Telecommunication Engineering, South China Normal University, Guangzhou 510006, China}
\affiliation{Guangdong-Hong Kong Joint Laboratory of Quantum Matter, Frontier Research Institute for Physics, South China Normal University, Guangzhou 510006, China}

\date{\today}
\begin{abstract}
Time synchronization and phase shaping of single photons both play fundamental roles in quantum information applications that rely on multi-photon quantum interference. Phase shaping typically requires separate modulators with extra insertion losses. Here, we develop a all-optical built-in phase modulator for single photons using a quantum memory. The fast phase modulation of a single photon in both step and linear manner are verified by observing the efficient quantum-memory-assisted Hong-Ou-Mandel interference between two single photons, where the anti-coalescence effect of bosonic photon pairs is demonstrated. The developed phase modulator may push forward the practical quantum information applications.
\end{abstract}
\pacs{42.50.Dv,03.67.Bg,42.65.Lm} 
\maketitle

\section{Introduction}
Multi-photon interference of quantum light, like single photons, lies at the heart of quantum information applications such as realizing quantum computing with linear optics and building large-scale quantum networks \cite{HOM,LQCRev2007,DLCZ2001,QRepeaterRMP2011,KimbleNature2008QNet,Wehner2018QNet}, which demands both the phase control and the timing sychronization operation of those photons. In one hand, phase degrees of those single photons not only play a fundamental role in manipulating the interference but also can be utilized to encode quantum information in quantum information applications protocols\cite{Lukin2020Nat,DPSQKD2002,TimebinQKD2004,Walmsley2020PS}. Hence, efficiently manipulating the phase of single photons is always desired for multi-photon interference. Separate optical elements, e.g., electro-optic phase modulator, are usually used to shape a single photon's phase\cite{phaseshape2009}, which might inevitabley reduce the multi-photon interference rate in the practical applications\cite{Lukin2020Nat}. On the hand, quantum memory (QM) that can store and readout singlel photons on-demand has been explored and demonstrated to be necessary for multi-photon interferometer\cite{PanRMP2012,ChenPRL2016} and large-scale quantum information applications\cite{QMrevTittel2013,NunnJ2013,Kwiat2017,Pan3node2019,Lukin2020Nat}. Therefore, developing a single device that can both synchronize single photons and simultaneously control their phases would push forward the practical quantum information applications. 

QM have been built based on various coherent quantum processes, such as electromagnetically induced transparency (EIT) \cite{EIT1991Harris,EIT2018Yu,LauratNC2018,Yan2019Qmem} and Raman schemes\cite{Ram2010Walmsley,Ram2015Guo,ZWProomraman2019}. The intrinsic coherence offers a promising way to tune the phase of light in the write and readout procedure\cite{DSP2000,patnaik2004pra,chen2015prl,qiu2016optica,QMbs2018}, which have also been implied by various experiments using weak laser pulses\cite{Lukin2002PRA,LukinRMP2003,Moon2017}. However, up to now, there have been no direct demonstration of phase control of quantum light such as a single photon using these schemes. 


In this work, using an efficient EIT-based QM, we firstly show a QM-assisted Hong-Ou-Mandel (HOM) interference by synchronizing the single photons spontaneously generated from two cold atomic ensembles, where the detectable interference event rate is improved by a factor of 15 comparing to the case without a QM. By phase-modulating the control laser of QM, we demonstrate an all-optical phase modulation of single photons while keeping the high storage efficiency of 86\%. Controllable coalescence and anti-coalescence behaviours of single photons are observed when the phase of readout single photons are phase modulated in both step and linear way. Our work verifies that both the coherence and quantum nature of single photons can be preserved well when modulated by a QM built-in phase modulator and may have promising applications in practical quantum information processing.

\begin{figure*}
\centering\includegraphics[width=15 cm]{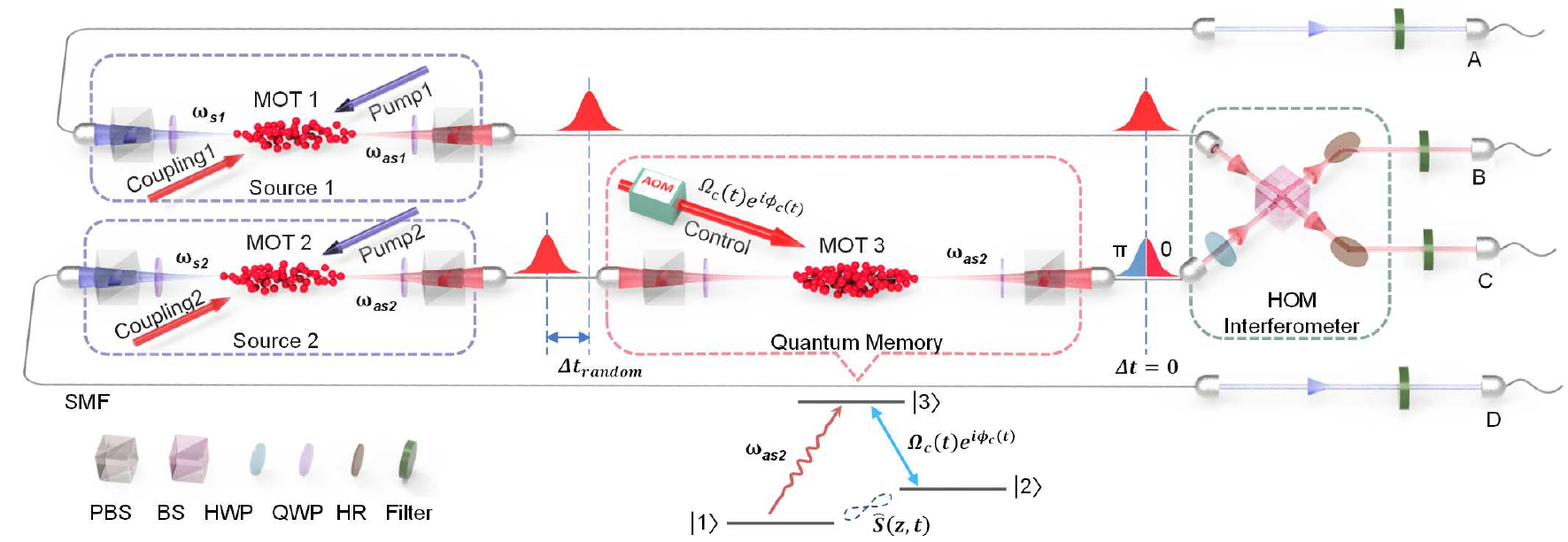}\\
\caption{\label{fig:sys} \textbf{Experimental setup.} Three cigar-shape dense cold atomic ensembles are prepared by dark-line magneto-optical traps (MOT) of $^{85}$Rb atoms. Single photons $\omega_{as1}$ ($\omega_{as2}$) heralded by its counterpart $\omega_{s1}$ ($\omega_{s2}$) are generated from MOT$_{1}$ (MOT$_{2}$) with the existence of pump1-coupling1 (pump2-coupling2) laser beams via the spontaneous four-wave-mixing process. Therefore, the timing differences between $\omega_{as1}$ and $\omega_{as2}$ are random and denoted by $\Delta t_{random}$. MOT$_{3}$ acts as an efficient QM based on EIT that can synchronize the readout single photon $\omega_{as2}$ to $\omega_{as1}$ ($\Delta t=0$). As shown by the energy level schematics of EIT two-photon process, a control laser manipulates the storage and readout of single photons $\omega_{as2}$. The amplitude and phase of the control laser pulse with a complex envelope of Rabi frequency $\Omega_{c}(t)e^{-i\phi_{c}(t)}$ is controlled by an acousto-optic modulator (AOM), by which the readout single photons $\omega_{as2}$ can be phase modulated accordingly. Single photons $\omega_{as1}$ and $\omega_{as2}$ are delivered to a HOM interferometer consisting of a beam splitter (BS) via single mode fibers (SMFs). Photons $\omega_{s1}$ and $\omega_{s2}$ are also collected and sent to detectors via SMFs. The generated photons are eventually detected by single photon counting modules (SPCMs) A, B, C and D. Filters are inserted before SPCMs to filter out noisy photons. PBS:  polarization beam splitter, HWP: half wave plate, QWP:  quarter wave plate,and HR: high reflection mirror.}
\end{figure*}

\begin{figure*}
\centering\includegraphics[width=15 cm]{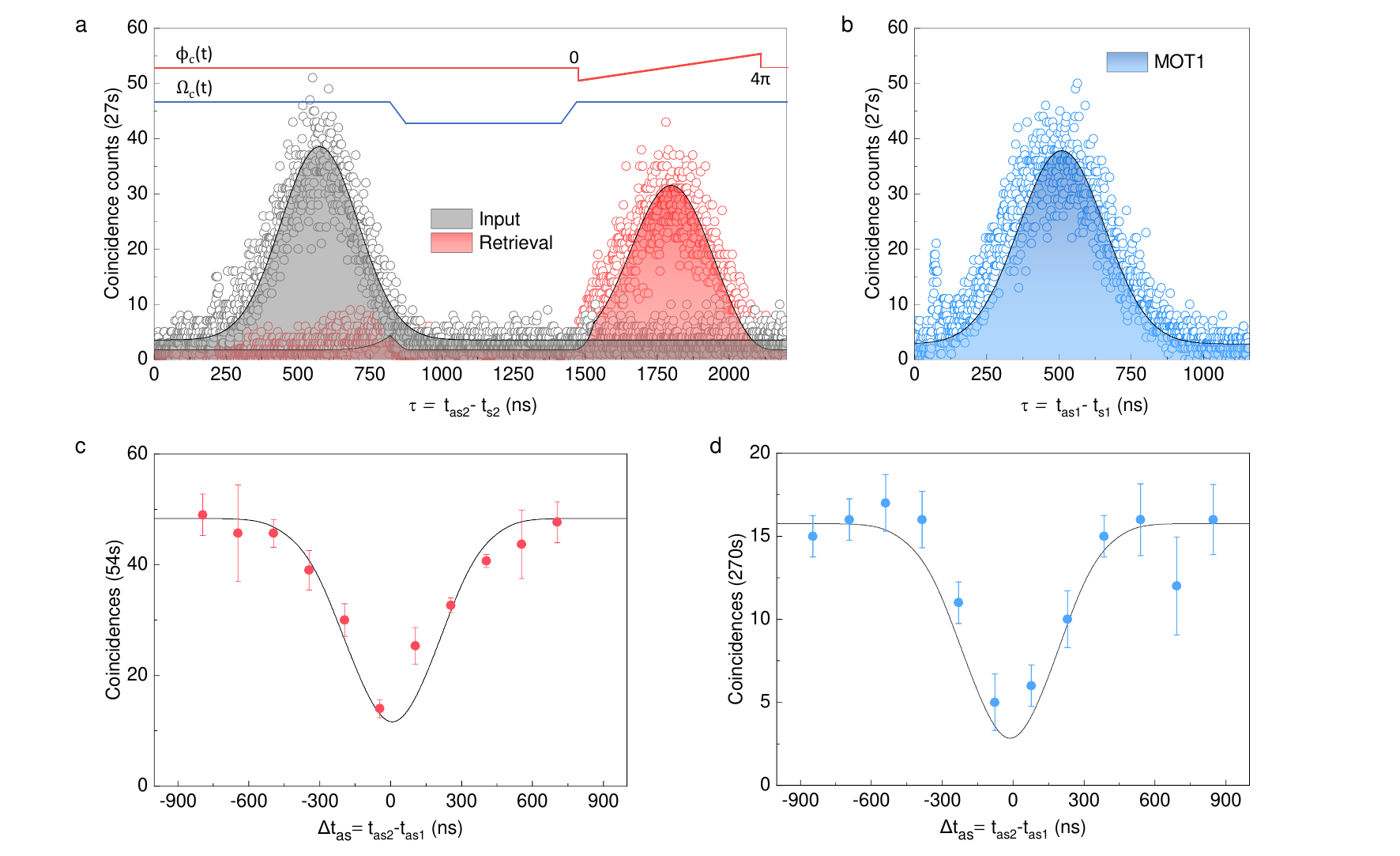}\\
\caption{\label{fig:HOM} \textbf{Synchronized HOM interference.}
\textbf{a.} Temporal waveforms of the single photons $\omega_{as2}$ (grey circles) and  the readout single photons (red circles). The blue solid line denotes the results of switching off the control laser with a storage time of 700~ns. The red solid line depicts the linear phase modulation of the control laser from 0 to 4$\pi$.
\textbf{b.} The temporal waveform of the single photons $\omega_{as1}$ emitted from MOT$_{1}$ depicted with blue circles. The solid lines enclosing shadow areas are Gaussian fitted guide lines of the above temporal waveforms.
$\tau$ is the detection time difference between  paired $\omega_{s1,2}$ and $\omega_{as1,2}$.
\textbf{c} and \textbf{d.} Synchronized and conventional HOM interference pattern measured by fourfold coincidence counts of $\omega_{s1}$, $\omega_{as1}$, $\omega_{s2}$ and $\omega_{as2}$ (See Appendix for details). Red dots and blue dots are experimental data points with error bars. Black solid lines are theoretical curves.}
\end{figure*}

\section{Theoretical model.}
The storage and readout procedure of a single photon in an EIT-base quantum memory (QM) can be well described by the dynamics of dark-state polariton\cite{DSP2000} inside an atomic ensemble including homogeneously distributed $N$ atoms:
\begin{eqnarray}\begin{array}{ll}
\hat\Psi (z,t) = cos\theta(t)\hat\psi(z,t) - sin\theta(t)\sqrt{N}\hat S(z,t)
\end{array}\end{eqnarray}
where $\hat\psi(z,t)$ refers to the spatial-temporal envelope of photon field inside the atomic ensemble and $\hat S(z,t)$ is the envelope of atomic spin wave. $tan\theta(t)=\frac{g\sqrt{N}}{\Omega_{c}(t)}$ with $\Omega_{c}(t)$ as the Rabi frequency of control laser and $g$ as the atom-photon coupling constant. Considering a typical storage procedure with initial constant $\Omega_{c}(t<0)=\Omega_{w}$, a single photon entering the atomic ensemble from the vacuum ($N=0$) would excite an atomic spin wave $\hat S(z,t)=\frac{g}{\Omega_{w}}\hat\psi(z,t)$. Then, the control laser is adiabatically switched off and kept off during $0\le t\le t_r$ to store the envelope profile of photon field into the stationary spin wave $\hat S(z,t)=\hat S(z,0)$ with zero group velocity $v_g=c_0*cos\theta(t)=0$ with $c_0$ as the light speed in vacuum (See Eq.(11) in Ref\cite{DSP2000}). 

When readout procedure starts, a control laser is switched on at $t=t_r$ with envelope of $\Omega_{r}(t-t_r)e^{-i \phi_{r}(t-t_r)}$. The propagating dark-state polariton $\hat\Psi (z,t)$ are adiabatically re-built up with $\hat S(z,t\ge t_r)=\frac{g\hat\psi_{r}(z,t-t_r)}{\Omega_{r}(t-t_r)e^{-i \phi_{r}(t-t_r)}}$, which eventually leave the atomic ensemble and is transformed into a readout photon field in vacuum. In the above ideal storage-readout process, the input and readout photon field are connected by $\hat S(z,t_r)=\hat S(z,0)$ with the following relation, except an global undetermined dynamical phase factor $e^{-i\phi_{s0}}$ might be acquired in experiment due to control laser phase fluctuation, atomic spin wave dynamics and so on:
\begin{eqnarray}\begin{array}{ll}
\hat\psi_{r}(z,t-t_r)=\hat\psi(z,t)\frac{\Omega_{r}(t_{r})e^{-i\phi_{r}(t_{r})}}{\Omega_{w}e^{i\phi_{s0}}}
\end{array}\end{eqnarray}

\section{Results}
The experimental setup shown in Fig.~\ref{fig:sys} contains three magneto-optical traps (MOT$_1$, MOT$_2$ and MOT$_3$ ) of $^{85}$Rb atoms and a HOM interferometer. Cold atomic ensembles in MOT$_{1}$ and MOT$_{2}$ serve as two independent single photon sources. The dense cold atomic ensemble in MOT$_{3}$ serves as an efficient QM. Heralded single photons $\omega_{as1}$ and $\omega_{as2}$ are spontaneously and independently generated from MOT$_{1}$ and MOT$_{2}$, respectively. $\omega_{as2}$ can be stored in the QM immediately after its generation which is triggered by detecting a $\omega_{s2}$ photon. When a photon $\omega_{s1}$ is detected at the instant $t_{s1}$, the stored single photon $\omega_{as2}$ can be readout with a controllable delay respective to $t_{s1}$ and then enter the HOM interferometer together with the single photon $\omega_{as1}$ that is also heralded by $\omega_{s1}$.

In the experiments, single photons $\omega_{as1}$ and $\omega_{as2}$ are made almost indistinguishable by carefully tuning their polarization states, spatial modes, temporal and spectral profiles(See Appendix for details).  Single photons $\omega_{as1}$ and $\omega_{as2}$ are experimentally generated in the same spontaneous four-wave-mixing processes that ensures they have the same polarization states and temporal and spectral profiles, where the optical depth (OD), the intensity and polarization of pump-coupling lasers and the focal point position of the pump laser are all carefully tuned in the photon sources of MOT$_1$ and MOT$_2$\cite{LiJF2019}. 
Their spatial modes before entering HOM interferometer are shaped the same by firstly being coupled into and then released out of the same type of singe mode fibers (SMFs). 
As shown in Fig.~\ref{fig:HOM}a and b, the Gaussian shape temporal waveform likeness between $\omega_{as1}$ and $\omega_{as2}$ is 98.2\% with a full width of half maximum of 320 ns. Meanwhile, the quantum nature of generated single photons are characterized by measuring their conditional auto-correlation function $g_{c}^{(2)}$ using the standard Hanbury Brown–Twiss interferometer. We have $g_{c}^{(2)}$ of 0.34 for both $\omega_{as1}$ and $\omega_{as2}$ photons, which verify their single photon quality.

\begin{figure*}
\centering\includegraphics[width=15 cm]{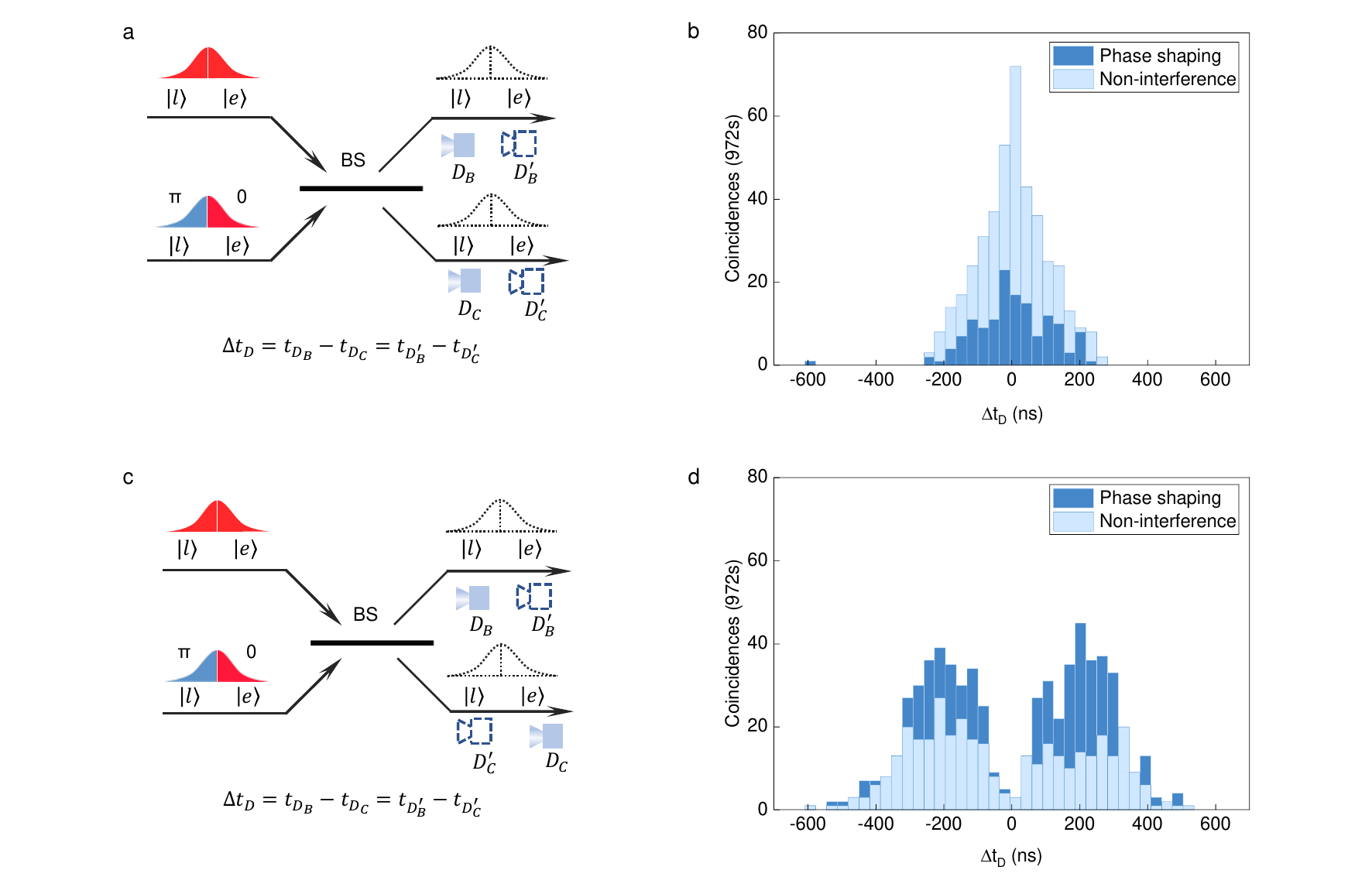}\\
\caption{\label{fig:shape} \textbf{Step phase modulation of $\pi$.}
\textbf{a,} and \textbf{c.} Schematics of detecting the two-photon wavepacket out of the HOM interferometer. $\Delta t_{D}$ is the detection time delay.
In \textbf{a} and \textbf{b.}, the two photons are detected in either the early ($|e,e\rangle$ at $t_{D_{B}}$ and $t_{D_{C}}$) or latter ($|l,l\rangle$ at $t_{D'_{B}}$ and $t_{D'_{C}}$)  half of the wavepacket.
In \textbf{c} and \textbf{d.}, the two photons are detected in different halves of the wavepackets by ($|l,e\rangle$ at $t_{D_{B}}$ and $t_{D_{C}}$) or ($|e,l\rangle$ at $t_{D'_{B}}$ and $t_{D'_{C}}$).
\textbf{b} and \textbf{d.} Two photon coincidences with time binwidth of 32~ns. The dark and light blue histograms show the interfered and non-interfered coincidence results, respectively. Owing to the finite rise time of the phase modulator, photons detected within $\pm25~ns$ are omitted from the above data.}
\end{figure*}

\begin{figure*}
\centering\includegraphics[width=15 cm]{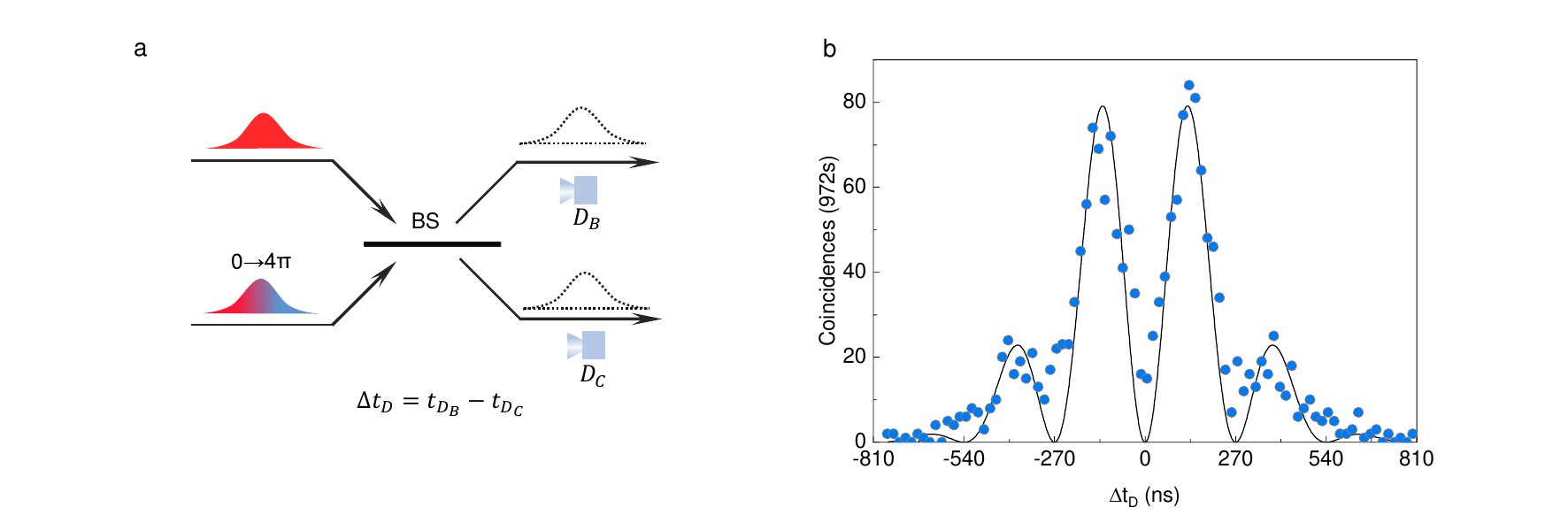}\\
\caption{\label{fig:4pi} \textbf{Linear phase modulation of $4\pi$.}
\textbf{a.} Schematic of detecting the two-photon wavepacket out of the HOM interferometer. $\Delta t_{D}$ is the detection time delay when a photon is detected by $D_{B}$ at time $t_{D_{B}}$ and another photon is detected by $D_{c}$ at time $t_{D_{c}}$.
\textbf{b.} Two photon coincidences with time binwidth of 18~ns. The blue dots show the experimental data points. The black solid line is the theoretical curve. }
\end{figure*}

\subsection{Synchronized HOM interference.}
To efficiently eliminate the arriving time difference at the HOM interferometer, a QM built in MOT$_{3}$ that is optimized for single photon $\omega_{as2}$ is inserted between MOT$_{2}$ and the HOM interferometer. The QM is made in cold atomic ensemble of $^{85}Rb$ atoms prepared in MOT$_{3}$\cite{Yan2019Qmem}. Atoms are pumped to Zeeman sublevel $|1\rangle=|5S_{1/2}, F=2, m_F=2\rangle$ before experimental window. Atomic OD 400 is measured on the transition $|1\rangle\leftrightarrow|3\rangle=|5P_{1/2},F=3, m_F=3\rangle$ for circularly polarized anti-Stokes field. The control laser is tuned to be resonant transition between $|3\rangle\leftrightarrow|2\rangle=|5S_{1/2},F=3, m_F=2\rangle$. The single photon field is focused at the center of MOT$_3$ with a beam waist radius of 150~$\mu$m. The control laser shines on atomic cloud with an angle of $1^{o}$ and a $1/e^2$ diameter 4.7~mm. As shown in Fig.~\ref{fig:HOM}a, a high storage efficiency of 86\% is achieved in our QM. Additionally, the temporal waveform likeness (98.5\%) of photon $\omega_{as2}$ together with its polarization state, spatial modes, frequency and single-photon nature ($g_{c}^{(2)}$=0.43) are all well-preserved during the storage-readout process.

HOM interference with and without QM are experimentally compared in our work. In Fig.~\ref{fig:HOM}c, the QM assisted HOM interference shows a HOM visibility of 76\%. The horizontal axis denotes the arrival time delay $\Delta t_{as}$ between single photons $\omega_{as1}$ and $\omega_{as2}$, which is swept by controlling the storage time of $\omega_{as2}$ with a step of 150~ns. The coincidence counts are measured at the exit ports of the HOM interferometer with an integration window of 640~ns that covers the whole single-photon wavepacket.  The results from conventional HOM interference are plotted in Fig.~\ref{fig:HOM}d, where single photons $\omega_{as1}$ and $\omega_{as2}$ enter the HOM interferometer randomly. The HOM interference pattern is obtained by passively analyzing the fourfold coincidence events with respect to their arrival time difference $\Delta t_{as}=t_{as2} -t_{as1}$ for QM assisted HOM and $\Delta t_{as}=t_{s2}-t_{s1}$ for HOM without a QM. The visibility of HOM dip measured in this way is 82\%, which is mainly limited to the finite distinguishability of photons from the source. Using the experimental data shown in Fig.~\ref{fig:HOM}c and d, the measured coincidence rates are $0.89s^{-1}$ and $0.06s^{-1}$, respectively.
Therefore, the detected four photon coincidence rate of synchronized the HOM interference is improved by a factor of around 15  compared to that of the conventional HOM interference. The coincidence rate is limited by the finite storage time of 5~$\mu s$ in our work and can be further improved if the storage time is longer.

\subsection{Phase modulation using quantum memory.}
From Eq.(1), by modulating the phase of control laser $\phi_{r}(t-t_{r})$ during the readout process, the readout single photon might be phase modulated in the same way. Although the amplitude envelope of the readout single photons can also be modulated in the same way, in our current work, the amplitude of control laser envelopes is optimized for keeping a high storage efficiency\cite{Yan2019Qmem}. We firstly realize a step phase change of $\pi$ of the readout single photon and verify it in our HOM interferometer. In priciple, the different phase envelopes make two input photons completely distinguishable and would destroy the HOM dip when the whole single photon wavepackets are take into account, which means that all the interference event are simply counted. However, when observing the HOM interference of the partial wavepacket of two single photons, we may still obtain interesting HOM interference pattern\cite{phaseshape2009}. We thus investigate the HOM interference pattern in the time domain by separating the photon wavepacket into two halves ($|e\rangle$  is the earlier half, and $|l\rangle$ is the later half).

When photon coincidence is detected in the same halves $|e,e\rangle$ or $|l,l\rangle$, as illustrated in Fig.~\ref{fig:shape}a, the photons behave in a coalescence way and tend to exit at same out port of BS as expected, because the two input states $|e,e\rangle$ and $|l,l\rangle$ are temporally separated and thus each input paired half can be viewed as regular HOM interference. The interfered two-photon wavepacket is shown in Fig.~\ref{fig:shape}b, where the coincidence count is significantly reduced to only 30\% of that of the non-interference case. The non-interference coincidence count is measured by intentionally adjusting the polarization states of single photons $\omega_{as1}$ and $\omega_{as2}$ to be orthogonal to each other and thus making them distinguishable. However, when photon coincidence is detected in different halves of the wavepacket $|e,l\rangle$ (or $|l,e\rangle$), as illustrated in Fig.~\ref{fig:shape}c, photons behave in an anti-coalescence way. Take the detected output state $|e,l\rangle$ as an example, the reflected input state of $|e,-l\rangle\rightarrow|e,l\rangle$ and the transmitted input state $|e,l\rangle\rightarrow|e,l\rangle$ constructively interfere at the exit ports of BS. When the phase of the read-out single photon is modulated, the interfered two-photon wavepacket shown in Fig.~\ref{fig:shape}d reveals that the coincidence count increases up to 170\% compared to the non-interference case. Our results verify that a step phase modulation of $\pi$ of the readout single photon $\omega_{as2}$ via QM is realized, which agrees well with the theoretical prediction of coincidence count changes of 0\% and 200\%.

In addition, we demonstrate a linear phase modulation of $4\pi$ spanned over the whole single photon wavepacket. As shown in Fig.~\ref{fig:4pi}, with a linear phase modulation within 540 ns, the expected oscillation structure in the interfered two-photon wavepacket can be clearly observed. The experimental data points (blue dots) agree well with the theoretical curves (black solid line) with an oscillation period of 270 ns, which is the time for generating $2\pi$ phase modulation (See methods). Alternatively, this oscillation structure is  the beating signal in two-photon interference when two input photons have a finite center frequency difference introduced by linear phase modulation\cite{Kuhn2004PRL}. Therefore, the above results clearly show that arbitrary phase modulation can be realized together with the storage-readout processes of an EIT-based QM while the storage efficiencies are kept above 86\%. 

\section{Conclusion and outlook}
In conclusion, we developed a phase modulator built in an EIT-based atomic QM. Synchronized HOM interference of spontaneously generated single photons is demonstrated with the assistance of a QM.  The interference rate increases  by a factor of 15  compared to the conventional case based on post-selection. Furthermore, the phase modulation of the readout single photon wavepacket out of QM is demonstrated while  quantum storage performance remains high. Both step phase modulation and linear phase modulation of the read-out single photon are realized by directly modulating the phase envelope of the control laser. The coalescence and anti-coalescence behaviours in the HOM interference are explored by observing the two-photon wavepacket out of the interferometer, which successfully verifies the expected phase modulation of single photons.

Our single photon phase modulator causes  no additional photon loss and thus would be of broad interest to multiple research areas in quantum information processing, such as memory required quantum communication\cite{Yantimebin2011,Pan3node2019,Lukin2020Nat,Wehner2018QNet},  quantum information processing using temporal modes of single photons\cite{Walmsley2020PS}, and quantum computation using linear optics\cite{KnillNature2001}.

\section{Acknowledgements}
We thank both Prof. Peng Xu from Wuhan Institute of Physics and Mathematics of the Chinese Academy of Sciences and Prof. Jiefei Chen from Southern University of Science and Technology for their useful discussions. 

\section{Appendix}
\subsection{Heralded single photons sources.}
Two single photon sources with exactly the same configuration are built based on MOT$_{1}$ and MOT$_{2}$ of $^{85}Rb$ atoms. We take MOT$_{1}$ as an example to illustrate the configuration. Laser cooled atoms are optically pumped to the lowest hyperfine level $|5S_{1/2}, F=2\rangle$ before experimental window. Optical depth of atomic cloud is 120 on the transition $|5S_{1/2}, F=2\rangle\rightarrow|5P_{1/2}, F=3\rangle$, measured with a circularly polarized probe field. A pair of  counter-propagating pumping laser beams pump$_{1}$ (780 nm, $\sigma^{-}$) and coupling laser beams coupling$_{1}$ (795 nm, $\sigma^{+}$) are shone on MOT$_{1}$ with an angle of $2.75^{o}$ to the quantization axis. Counter-propagating entangled photon pairs are generated along the quantization axis. One of the photons, Stokes photon $\omega_{s1}$ is detected by single-photon counting modules (SPCM, Excelitas, SPCM-AQRH-16), which heralds the generation an anti-Stokes photon $\omega_{as1}$. The pump laser is blue detuned from the transition $|5S_{1/2},F=2\rangle\leftrightarrow|5P_{3/2},F=3\rangle$ by $150MHz$ and focused on the center with a  $0.82 mm$ $1/e^2$ diameter and power of 14$\mu W$. The coupling laser, resonant to the transition $|5S_{1/2},F=3\rangle\leftrightarrow|5P_{1/2},F=3\rangle$, is collimated Gaussian beam with $1/e^2$ diameter of $1.8 mm$ and power of $3mW$.The optimal Gaussian-like single-photon temporal waveform is obtained  following the optimization procedure described in Refs\cite{LiJF2019}. Configuration of pump2 and coupling2 has slightly difference in power and beam side. Laser pump2 has focal point spot diameter of $0.95 mm$ and power of $12.5\mu W$. Laser coupling2 is collimated with diameter of $1.8 mm$ and power of $4.5 mW$. After detected by SPCMs, coincidence counts between photons are recorded by a time-to-digital converter (Fast Comtec P7888) with a time bandwidth of 2 ns. The experiment runs repeatedly with a rate of 50~Hz including an experimental window of 300~$\mu s$.

\subsection{Fourfold coincidence count.}
The fourfold coincidence is acquired by analyzing the counting relation of $\omega_{s1}$, $\omega_{s2}$, $\omega_{as1}$ and $\omega_{as2}$ photons. After HOM interferometer, the fourfold coincidence rate\cite{ChenPRL2016}:
\begin{equation}
\begin{aligned}
P^{(4)}=&T^{2}\int d\tau|\psi_{1}(\tau)|^{2}+R^{2}\int d\tau|\psi_{2}(\tau)|^{2}\\
&-2TR|\int d\tau\psi^{*}_{1}(\tau)\psi_{2}(\tau+\triangle t)|^{2}\\
&+TR\tilde{g} ^{(2)}\int d\tau|\psi_{1}(\tau)|^{2}\int d\tau|\psi_{2}(\tau)|^{2},
\end{aligned}
\end{equation}
where $\psi_{1}(\tau)$ ($\psi_{2}(\tau)$)is the single photon wave packet function of MOT$_1$ (MOT$_2$), and $\tau$ is relative delay between $\omega_{as1}$ and $\omega_{as2}$. The $\triangle t$ is time difference of $\omega_{s1}$ and $\omega_{s2}$. T and R refer to the transmission and reflection of BS, respectively. The $\tilde{g} ^{(2)}$ is the average value of ${g} ^{(2)}$ within whole wave packet.

\subsection{Two photon coincidence after HOM interferometer.}
When dynamical varying the phase of one photon, the photon coincidence between two BS output port becomes phase-dependent. The coincidence rate\cite{phaseshape2009}:
\begin{eqnarray}
\begin{array}{ll}
&P^{2}= |\psi_{1}(\tau)|^{2}|\psi_{1}(\tau)|^{2}\sin(\pi\triangle\nu\tau),
\end{array}
\end{eqnarray}
where the $\triangle\nu$ is phase-varying-induced frequency difference. In our linear phase shaping result, $4\pi$ variation within 540 ns, the phase-varying-induced frequency difference is 3.7 MHz.

\bibliographystyle{ieeetr}
\bibliography{QMemHOMbib}

\end{document}